\documentclass[twocolumn,showpacs,preprintnumbers,amsmath,aps,amssymb]{revtex4-1}
\usepackage{graphicx}
\usepackage{dcolumn}
\usepackage{bm}
\usepackage{color}

\begin{document}

\title{A general procedure to find ground state solutions for finite $N$ M(atrix) theory. Reduced models and SUSY quantum cosmology}

\author{J. L. L\'opez}
\email{jl\_lopez@fisica.ugto.mx}
\author{O. Obreg\'on}
\email{octavio@fisica.ugto.mx}
\affiliation{ Departamento  de F\'{\i}sica de la Universidad de Guanajuato,\\
 A.P. E-143, C.P. 37150, Le\'on, Guanajuato, M\'exico. \\
 }
\begin{abstract}
We propose a general method to find exact ground state solutions to the SU($N$) invariant matrix model arising from the quantization of the 11-dimensional supermembrane 
action in the light-cone gauge. We illustrate the method by applying it to lower dimensional models and for the SU(2) group. This approach can be used to find ground state solutions to 
the complete 9-dimensional model and for any SU($N$) group. The supercharges and the constraints related to the SU($2$) symmetry are the relevant operators and they generate a multicomponent  
wave function. In the procedure, the fermionic degrees of freedom are represented by means of Dirac-like gamma matrices.  
We exhibit a relation between these finite $N$ matrix theory ground state solutions and SUSY quantum cosmology wave functions giving a possible physical significance to the theory even for finite $N$\end{abstract}
\pacs{11.25.-w, 04.60.Ds, 04.65.+e, 04.60.Kz, 12.60.Jv}
\maketitle

\section{Introduction}

As is known, M-theory should be an eleven dimensional supersymmetric quantum theory from which all superstring theories can be deduced and it can be connected also to 11-dimensional supergravity. 
This idea was born since it was realized that the five consistent theories of strings and 11-$d$ supergravity are related among them through several types of dualities \cite{1.1,1.2,1.3,1.4,1.5}. 
There are interesting approaches and calculations that provide the possibility to understand some important features of this fundamental theory. There is a concrete formulation of M-theory which 
has been claimed has the appropriate expected properties and it is based on the matrix model conjectured in (BFSS \cite{1.1}). This matrix model is a supersymmetric quantum mechanics where the bosonic 
degrees of freedom are a finite set of $N \times N$ Hermitian matrices \cite{Sato}. We can associate a time dependent matrix to every point in space and these matricial coordinates are obviously noncommuting. 
The existence of dynamical objects of dimensionalities higher than one, like $Dp$-branes, has left aside the fundamental status of particles and strings and the origin of this matrix model and its connection to 
M-theory arises from the dynamics of D0 branes \cite{Polchinski,Witten,Danielsson}. It was conjectured in \cite{1.1} that M-theory is the $N \rightarrow \infty$ limit of the maximally supersymmetric matrix Hamiltonian 
emerging from the D0-brane matrix model. The conjecture was extended in \cite{1.6} showing that matrix theory is also meaningful for finite $N$, so the solutions to the matrix model for $N$ finite are relevant.

In the recent years, research has been done on another matrix model arising from noncommutative gauge theories \cite{ikkt1}, it is called the IKKT matrix model and it can also be related to a large $N$ reduced 
model of 10-dimensional non-commutative Super Yang-Mills theory. In the IKKT framework, the parameter $\theta$ that measures the non-commutativity between the coordinates of space-time becomes dynamical 
and in the semiclassical limit it defines a Poisson structure such that the metric of space-time itself emerges naturally \cite{stkr1,stkr2,stkr3}. This matrix model can be seen as a $\mathcal{N} = 4$ noncommutative 
Super Yang-Mills theory on $\mathbb{R}^4_{\theta}$. The path integral quantization of this matrix model is naturally defined by integrating over the space of matrices and in a semiclassical way, it reproduces some 
interesting features of curved spacetime \cite{stkr4}. So, it has been claimed is a good candidate to be a quantum theory of gravity together with other fundamental interactions \cite{stkr5}. The presence of branes in this 
theory \cite{stkr6} becomes also important as it is in the matrix theory of BFSS. The quantization approaches are different for the two matrix models briefly described above, but they have the same common D$p$-brane 
world origin and both involve the most important characteristic features of a unified theory of physical interactions including gravity.
 
Another, and somehow independent, matrix model was encountered in the quantization of the classical 11-dimensional supermembrane action in the light cone gauge (LCG) \cite{N1,N2,N3,Halpern1}. 
These matrix models were studied in the context of supersymmetric quantum mechanics years before the quantization of the supermembrane was finally achieved \cite{Halpern}. The action of the supermembrane is the 
generalization of the Green-Schwarz action for the superstring for a membrane moving in a $d$-dimensional target superspace, and when $d= 11$ this can be directly related to 11-dimensional supergravity \cite{Sezgin}. 
It is possible to regularize the theory because of the existence of a finite dimensional Hilbert space where the area preserving diffeomorphisms (APD) invariant states can be expressed. This APD invariance is 
translated into a SU($N$) symmetry in the regularized theory which is equivalent to a Super Yang-Mills quantum mechanics matrix model \cite{N1,N2,N3}. The Hamiltonian and the associated ground state of these 
matrix models of supersymmetric quantum mechanics with SU($N$) invariance were studied in \cite{Halpern} and the matrix model arising from the supermembrane action is exactly one of those. In this type of 
supersymmetric models it is of great importance to find the ground state solution \cite{Halpern,Halpern1,Sethi1,Sethi2}, and for the matrix model we have been briefly mentioned here the existence of such state, even for $N$ finite, has been found only perturbatively.

These two matrix models, the one related to D0-branes and the other related to the supermembrane, are not independent. There is a connection between membranes and D0-branes first speculated 
in \cite{Town} where it was considered that membranes could be regarded as collective excitations of D0-branes. This connection was treated in great detail in \cite{1.1}. 
The noncommutative space defined by the phase space coordinates where the membranes exist has a basic minimum area, this quantum unit cells of space are precisely the D0-branes, each one 
with a longitudinal momentum of 1/$N$. The longitudinal direction is the one compactified in the LCG description. This relation of duality connects the two matrix models. 

The first purpose of this work is to exemplify the general method by finding solutions to  the ground state of  matrix theory reduced models arising from the quantization of the classical supermembrane action. 
Our procedure is based on; i) We consider finite $N$ matrix models, ii) The relevant operators are the supercharges and those related with the SU($N$) symmetry. It is no more necessary to consider 
the Hamiltonian operator because through the canonical algebra it is a consequence of these constraints, iii) The fermionic degrees of freedom are represented utilizing a Dirac-like gamma 
matrix representation \cite{jefe1,jefe4}. By means of these, the associated supercharges and the operators related to the SU($N$) symmetry become matrix differential operators acting on a multicomponent 
wave function. One generates a Dirac-like quantum matrix model formalism.

With our proposal one can solve any finite $N$ matrix model and for all the bosonic and fermionic degrees of freedom. Even though finite $N$ matrix models are physically relevant \cite{1.6}, it is not clear 
which particular $N$ should be considered. For large $N$, our method will generate a large number of first order coupled differential equations. In order to illustrate how to apply our proposal, we restrict 
ourselves to reduced models, for the SU($N = 2$) invariance group. Besides, for the sake of simplicity and as a way to demonstrate how this procedure works, we will make further assumptions in the 
space of bosonic variables. We will then freeze out some bosonic degrees of freedom. By doing this one cannot assure that this reduced models will capture 
all features of the quantization of finite $N$ matrix theory (``for a related discussion in the context of quantum cosmology see; \cite{Ryan}''). 
Some of the models studied can however be exactly solved illustrating us about the kind of ground state solutions 
to be expected for a larger $N$ model. The method we propose can be directly extended to higher dimensional models and for any SU($N$) group. 
This kind of representation for the fermionic variables has been used in a first and several models of SUSY quantum cosmology \cite{jefe1,jefe4,jefe3,jefe6,Torres,SFmatter3} leading to 
pretty much interesting features regarding every physical supersymmetric system treated in this way. The fermionic degrees of freedom can also be represented by differential operators \cite{jefe2,DEath1,Moniz,PD2}. 
A relevant point is that it is possible to solve the complete operators system containing 
both, the bosonic and the fermionic sectors. The fermionic information will then be encoded in each one of the entries of the multicomponent wave function and the physical information 
regarding the variables of the system will be given by means of the components of this wave function.

It has been argued that M-theory is a promising candidate to be a quantum theory of gravity, it is then expected that the matrix models related with it, contain quantum and classical gravitational information. The same 
authors of the conjecture have shown that a specific compactification of matrix theory correctly describes the properties of a Schwarzschild black hole \cite{16.1,16.2,16.3,16.4} and other works related to matrix theory 
black holes and its thermodynamic properties are \cite{Bht1,Bht2,Bht3,Bht4}. 

It has also been claimed that classical cosmological models can be deduced directly from the matrix theory of BFSS \cite{Alvarez,Gibbons} 
and specifically in \cite{Gibbons}, using a homothetic ansatz, the authors were able to derive the Friedmann equations from the bosonic sector of the classical Hamiltonian of matrix theory. 
Following this kind of ideas, the interesting point to us here is that we show that some supersymmetric 
cosmological models are then hidden in the SU($N$) invariant supersymmetric matrix model as well. It is then expected to find supersymmetric quantum solutions that can be interpreted and related to SUSY quantum 
cosmology models. This is a particularly interesting byproduct of our reduced matrix models and shows that these reduced models are also supersymmetric given that their corresponding SUSY cosmological models 
are supersymmetric by construction \cite{jefe6,Torres}. Also recently \cite{Damour}, to describe the quantum dynamics of the supersymmetric Bianchi IX cosmological model \cite{jefe4} it has been proposed to fix 
from the start the six degrees of freedom describing local Lorentz rotations of the tetrad. The operational content of the supercharges and Hamiltonian 
revealed a hidden hyperbolic Kac-Moody structure which seems to support the conjecture about a correspondence between supergravity and the dynamics of a spinning particle moving on an infinite 
dimensional coset space. At the same time we show that some reduced matrix models correspond to exact SUSY quantum cosmology models that arise from supergravity.   
All these relations point out to a web of connections between these different topics and make it of interest to search for a method to find ground state solutions in finite $N$ matrix theory models. 
Here, we present a general procedure to construct them. We will also be interested in relations between the solutions of these ground state matrix theory models 
and wave functions corresponding to SUSY quantum cosmology. One of the wave function solutions obtained in 
our reduced models resembles a wave function solution obtained in the superfield approach to supersymmetric quantum cosmology and it is then argued that the matrix model we consider provides the relevant physical 
information, as classical cosmology seems also to arise from bosonic matrix models \cite{Alvarez,Gibbons}.

The work is organized as follows. First in section II we show the matrix model and its description as a supersymmetric SU($N$) invariant quantum mechanical model. In section III we illustrate the way in which we look for 
solutions with the matricial representation for the fermionic degrees of freedom in a 2-dimensional reduced model and for the SU(2) group. In section IV we show the solutions we obtain for some particular assumptions on 
the bosonic variables. As we pointed out, this method could be directly extended to the complete 9-dimensional theory and for any SU($N$) group, although for large $N$ powerful computational 
tools would be needed. 
Then in section V we make an extension to a 4 dimensional model. It is interesting to remark that some of the 2-dimensional solutions are also solutions to the 4-dimensional model. In section VI we show first that in this 
extension we encounter, exactly, a model guessed in \cite{N3} and used to analyze the spectrum of the matrix 
Hamiltonian of the quantum supermembrane action and then we exhibit some solutions to this extended model. In section VII we discuss a connection of one of the solutions to the extended 4-dimensional model with 
SUSY quantum cosmology. Section VIII is devoted to conclusions.          

\section{Matrix model}

As we mentioned in the introduction, the model arising from the quantization of the classical supermembrane action, once the regularization is achieved, becomes a model of supersymmetric quantum mechanics \cite{1.1}
 that can also be obtained from a dimensional reduction of a super Yang-Mills theory with gauge group SU($N$) \cite{Halpern,N1}, the Hamiltonian of this matrix model \cite{Halpern1} is given by  

\begin{align}
H & = \frac{1}{2}\pi_a^m \pi_a^m + \frac{g^2}{4}f_{abc} \phi^m_b \phi^n_cf_{ade}\phi^m_d \phi^n_e \label{Ham} \\ \nonumber
& ~~ -\frac{ig \hbar}{2} f_{abc}\Lambda_{a \alpha} (\Gamma^m)_{\alpha \beta} \phi^m_b \Lambda_{c \beta},  \\ \nonumber
& m,n = \{\ 1,2,..., 9 \}\ , ~~~a,b,c = \{\ 1,2,..., N^2-1 \}\ , \\ \nonumber
& \alpha, \beta = \{\ 1,2...,\mathcal{N} \}\ , 
\end{align}

\noindent where $\phi^m_a$ and $\pi^m_a$ are the bosonic degrees of freedom and its related momenta respectively, and $\phi^m_a$ are the components of the SU($N$) valued matrices $\Phi^m$ given by 
$\Phi^m = \phi^m_a \tau^a$ where $\tau^a$ are the basis  group elements. Then $a$ is a group index and $m$ is a dimension index. $\Lambda_{a \alpha}$ represent the 
fermonic degrees of freedom and $\Gamma^m$ are $m$ Dirac matrices obeying the Clifford algebra $\{\ \Gamma^m,\Gamma^n \}\ = 2\delta_{mn}$. The SU($N$) group structure constants are $f_{abc}$. 
The algebra between the supercharges as function of the total Hamiltonian and the supercharges themselves are given by 

\begin{align}
& \{\ Q_\alpha,Q_\beta \}\ = 2\delta_{\alpha \beta}H + 2g(\Gamma^n)_{\alpha \beta}\phi_a^nG_a , \label{salgebra} \\ \nonumber 
& Q_\alpha = (\Gamma^m \Lambda_a)_\alpha \pi^m_a + igf_{abc}(\Sigma^{mn} \Lambda_a)_\alpha \phi_b^m \phi_c^n,
\end{align}

\noindent where $\Sigma^{mn} = -\frac{i}{4}[\Gamma^m,\Gamma^n]$. In the anticommutator of the supercharges it appears the operator (constraint) related to the SU($N$) invariance $G_a$ which in terms of the bosonic 
and fermionic variables is given by 

\begin{equation}
 G_a=f_{abc}(\phi_b^m \pi_c^m - \frac{i \hbar}{2}\Lambda_{b \alpha} \Lambda_{c \alpha}).
\end{equation} 

\noindent Now, we want to realize the quantization. And as we are working with a supersymmetric quantum mechanics where the Hamiltonian is the square of the supercharges and it is positive semi-definite, the state 
$| \Psi \rangle$ which obeys $H |\Psi \rangle = 0$ is automatically the ground state wave function. Also, the ground state should be a gauge invariant state, it should satisfy $G_a |\Psi \rangle = 0$. The commutator and anticommutator relations  for the bosonic and fermionic degrees of freedom will be

\begin{equation}
[\phi_a^m,\pi_b^n] = i\hbar \delta_{ab} \delta_{mn},~~~~~~ \{\ \Lambda_{a \alpha},\Lambda_{b \beta} \}\ = \delta_{ab}\delta_{\alpha \beta}. \label{algebra}
\end{equation}

The supercharges arise as the square root of the Hamiltonian, and as it can be seen in the algebra eq. (\ref{salgebra}), we can solve the equations $Q_\alpha | \Psi \rangle = 0$ together with 
$G_a | \Psi \rangle = 0$. Then the ground state satisfying this, will respect all the symmetries; gauge and exact supersymmetry. The representation of the canonical momenta corresponding to $\phi^m_a$ is, as usual,
$\pi^m_a = -i\hbar \frac{\partial}{\partial \phi^m_a}$, and given the algebra for the fermionic variables eq. (\ref{algebra}), we have the freedom to represent the $\Lambda_{a \alpha}$ as matrices. 
This kind of representation was used first in SUSY quantum cosmology \cite{jefe1}. This is an essential tool \cite{jefe1,jefe3,jefe4,jefe6} in this work and as we will show, we will be able to find exact solutions 
to some reduced lower dimensional models. The method we will present can be applied to search for solutions to the 9-dimensional supersymmetric matrix model and for any SU($N$) group. 
However, as it will be seen, for these dimensions and a large $N$ this is a difficult task because one would need to solve an extremely large number of coupled partial differential equations.      

\section{The method of solution}

To begin with, we will try to find solutions to the ground state by solving the corresponding equations $Q_\alpha | \Psi \rangle = 0$ and $G_a | \Psi \rangle = 0$ (the Hamiltonian constraint is then satisfied) for a SU($2$) symmetry;  
$a = 1,2,3$ and for a 2-dimensional reduced model, $m = 1,2$. Hence, we will have six $\phi^m_a$ bosonic degrees of freedom and the same number for the fermionic ones. In the variables $\Lambda_{a \alpha}$,  $a$ 
is a group index  and the $\alpha$ index represents the number of supercharges which is in accordance with the number of components of a spinor in $d$-dimensions. Then with $d = 2$, we have two supercharges 
and consequently six fermionic variables. The dimension $M$ of the square $M \times M$ matrix representation $\Lambda_{a \alpha}$ depends on the number of fermionic degrees of freedom. 
So, in this first toy model $M = 8$. The representation we use for the $\Lambda_{a \alpha}$ matrices that satisfies eq. (\ref{algebra}) is the following  

\begin{align}
& \Lambda_{11} = \frac{\pm 1}{\sqrt{2}}\Delta^0, ~~\Lambda_{12} = \frac{ \pm i}{\sqrt{2}}\Delta^1, ~~\Lambda_{21} = \frac{\pm i}{\sqrt{2}}\Delta^2,  \label{rep2} \\ \nonumber
& \Lambda_{22} = \frac{\pm i}{\sqrt{2}}\Delta^3, ~~\Lambda_{31} = \frac{\pm i}{\sqrt{2}}\Delta^4, ~~\Lambda_{32} = \frac{\pm i}{\sqrt{2}}\Delta^5.  \\ \nonumber
& \Delta^{\mu} = \gamma^{\mu} \otimes \sigma^1, ~~~\Delta^4 = I \otimes i\sigma^2, ~\Delta^5 = I \otimes i\sigma^3, \\ \nonumber
& \mu = 0,1,2,3,
\end{align} 

\noindent where $\sigma^i$ are the Pauli Matrices, $\gamma^{\mu}$ are the Dirac matrices in a Majorana representation and $I$ is the $4 \times 4$ identity matrix. The algebra in eq. (\ref{algebra}) is not affected if we choose 
the plus, or minus, sign on every matrix in eq. (\ref{rep2}). Notice that we have three kinds of indices, those  corresponding to the supercharges $\alpha = 1,2$, to the  group $a = 1,2,3$, and the ones associated to the 
dimensions $m = 1,2$. Even when we have three kinds of indices we will denote all of them, as in eq. (\ref{rep2}), with the same kind of number. As an example, the first of the $\Lambda_{a \alpha}$ matrices is explicitly

\begin{align}
& \Lambda_{11} = \frac{\pm 1}{\sqrt{2}}\left( \begin{array}{ccccccccc} 
0 & ~0 & ~0 & ~0 & ~0 & ~0 & ~0 & -i \\
0 & ~0 & ~0 & ~0 & ~0 & ~0 & -i & ~0  \\
0 & ~0 & ~0 & ~0 & ~0 & ~i & ~0 & ~0  \\
0 & ~0 & ~0 & ~0 & ~i & ~0 & ~0 & ~0  \\
0 & ~0 & ~0 & -i & ~0 & ~0 & ~0 & ~0 \\
0 & ~0 & -i & ~0 & ~0 & ~0 & ~0 & ~0  \\
0 & ~i & ~0 & ~0 & ~0 & ~0 & ~0 & ~0  \\
i & ~0 & ~0 & ~0 & ~0 & ~0 & ~0 & ~0  
\end{array}
\right),
\end{align}

\noindent and for this model we choose the following two $\Gamma^m$ matrices 

\begin{equation}
\Gamma^1 =  \left( \begin{array}{ccc} \label{dirac2}
0 & ~1  \\
1 & ~0  \\  
\end{array} \right), 
\Gamma^2 =  \left( \begin{array}{ccc} 
1 & ~0  \\
0 & -1  \\  
\end{array} \right). 
\end{equation} 

The two supercharges $Q_{\alpha}$ of eq. (\ref{salgebra}) are given by the following operators

\begin{align}
Q_{1} & =  \Lambda_{11}\pi^2_{1}
 + \Lambda_{21}\pi^2_{2} + 
\Lambda_{31}\pi^2_{3} \\ \nonumber
& + \Lambda_{12}\pi^1_{1} + 
\Lambda_{22}\pi^1_{2} 
+ \Lambda_{32}\pi^1_{3} \\ \nonumber
& -g\Lambda_{12}(\phi^1_{2}\phi^2_{3} - \phi^1_{3} \phi^2_{2}) \\ \nonumber 
& - g\Lambda_{ 22} (\phi^1_{\bar{3}} \phi^2_{\bar{1}} - \phi^1_{\bar{1}} \phi^2_{\bar{3}}) \\ \nonumber
& - g\Lambda_{ 32}(\phi^1_{\bar{1}} \phi^2_{\bar{2}} - \phi^1_{\bar{2}} \phi^2_{\bar{1}}) ~, \\ \nonumber \\ \nonumber
Q_{2} & = \Lambda_{11}\pi^1_{1}
 + \Lambda_{21}\pi^1_{2} + 
\Lambda_{31}\pi^1_{3} \\ \nonumber
& - \Lambda_{12}\pi^2_{1} - 
\Lambda_{22}\pi^2_{2} 
- \Lambda_{32}\pi^2_{3} \\ \nonumber
& +g\Lambda_{11}(\phi^1_{2}\phi^2_{3} - \phi^1_{3} \phi^2_{2}) \\ \nonumber 
& + g\Lambda_{21} (\phi^1_{3} \phi^2_{1} - \phi^1_{1} \pi^2_{3}) \\ \nonumber
& + g\Lambda_{31}(\phi^1_{1} \phi^2_{2} - \phi^1_{2} \phi^2_{1}) ~, 
\end{align}

\noindent and the three operators $G_a ~(a = 1,2,3)$ related with the SU(2) gauge symmetry are given by 

\begin{align}
G_1  = & \phi^1_2 \pi^1_3 + \phi^2_2 \pi^2_3 - \phi^1_3 \pi^1_2 - \phi^2_3 \pi^2_2 \\ \nonumber
& - i \hbar (\Lambda_{21} \Lambda_{31} + \Lambda_{22} \Lambda_{32}), \\ \nonumber \\ \nonumber
G_2  = & \phi^1_3 \pi^1_1 + \phi^2_3 \pi^2_1 - \phi^1_1 \pi^1_3 - \phi^2_1 \pi^2_3 \\ \nonumber
& - i \hbar (\Lambda_{31} \Lambda_{11} + \Lambda_{32} \Lambda_{12}), \\ \nonumber  \\Ê\nonumber
G_3  = & \phi^1_1 \pi^1_2 + \phi^2_1 \pi^2_2 - \phi^1_2 \pi^1_1 - \phi^2_2 \pi^2_1 \\ \nonumber
& - i \hbar (\Lambda_{11} \Lambda_{21} + \Lambda_{12} \Lambda_{22}).
\end{align}

Using the matrix representation of eq. (\ref{rep2}), all of them are linear matrix differential operators acting on an eight component wave function $| \Psi \rangle$. We can restrict this model even more, setting 
some of the bosonic variables equal to zero, or making some identifications between them. These will give specific configurations. Using the anticommutator of the supercharges in eq.(\ref{salgebra}) and eq. (\ref{dirac2})
we can see that it is not always necessary to impose all of the three $G_a$ operators, and that we can solve the problem for one single operator $Q_\alpha$. For every 
$G_a |\Psi \rangle = 0$, and $Q_{\alpha} | \Psi \rangle = 0$, we have eight coupled partial differential equations for the eight components of the wave function $| \Psi \rangle = (\psi_1,\psi_2,\psi_3,\psi_4,\psi_5,\psi_6,\psi_7,\psi_8)$. 
For example the equation $Q_1 | \Psi \rangle = 0$ leads to the eight 
equations

\begin{widetext}
\begin{align}
& [3 f_1({\vec{\phi}}) - \pi^1_{3}  ] \psi_1
+ [  i\pi^2_{3} - \pi^1_1 +  3 f_2({\vec{\phi}}) ] \psi_2
+ [  \pi^1_{2}  -  3 f_3({\vec{\phi}}) ] \psi_4
- [  i\pi^2_{1} + \pi^2_{\bar{2}} ] \psi_8 = 0, \\ \nonumber
& [ 3 f_2({\vec{\phi}}) - i\pi^2_{3}  - \pi^1_1] \psi_1 
+ [  \pi^1_{3} - 3 f_1({\vec{\phi}}) ] \psi_2
+ [  \pi^1_{2} -  3 f_3({\vec{\phi}}) ] \psi_3 
- [  i\pi^2_{1} + \pi^2_2 ] \psi_7 = 0, \\ \nonumber
& [ \pi^1_{2} - 3 f_3({\vec{\phi}})] \psi_2
+ [  3 f_1({\vec{\phi}}) - \pi^1_{3} ] \psi_3
+ [  i\pi^2_{3}  + \pi^1_1 -  3 f_2({\vec{\phi}}) ] \psi_4 
+ [  i\pi^2_{1} + \pi^2_2 ] \psi_6 = 0, \\ \nonumber
& [ \pi^1_{2} - 3 f_3({\vec{\phi}})] \psi_1
+ [ \pi^1_1  - i\pi^2_{3} - 3 f_2({\vec{\phi}}) ] \psi_3
+ [  \pi^1_{3}  -  3 f_1({\vec{\phi}}) ] \psi_4 
+ [  i\pi^2_{1} + \pi^2_2 ] \psi_5 = 0, \\ \nonumber
& [ \pi^2_2 - i\pi^2_{1} ] \psi_4 
+ [ 3 f_1({\vec{\phi}}) - \pi^1_{3}] \psi_5 
+ [  i\pi^2_{3} - \pi^1_1 + 3 f_2({\vec{\phi}}) ] \psi_6
+ [  \pi^1_{2}  -  3 f_3({\vec{\phi}}) ] \psi_8 = 0, \\ \nonumber
& [  \pi^2_2 - i\pi^2_{1}] \psi_3 
- [  i\pi^2_{3} + \pi^1_1 - 3 f_2({\vec{\phi}})] \psi_5
+ [  \pi^1_{3} - 3 f_1({\vec{\phi}}) ] \psi_6
+ [  \pi^1_{2} -  3 f_3({\vec{\phi}}) ] \psi_7 = 0, \\ \nonumber
& [ i\pi^2_{1} - \pi^2_2] \psi_2 
+ [  \pi^1_{2} -  3 f_3({\vec{\phi}}) ] \psi_6
+ [ 3 f_1({\vec{\phi}}) - \pi^1_{3} ] \psi_7 
+ [ i\pi^2_{3}  + \pi^1_1 -  3 f_2({\vec{\phi}}) ] \psi_8 = 0, \\ \nonumber
& [ i\pi^2_{1} - \pi^2_2] \psi_1
+ [ \pi^1_{2} -  3 f_3({\vec{\phi}}) ] \psi_5
- [ i\pi^2_{3}  -  \pi^1_1 +  3 f_2({\vec{\phi}}) ] \psi_7
+ [ \pi^1_{3} - 3 f_1({\vec{\phi}}) ] \psi_8 = 0,
\end{align}
\end{widetext}

\noindent where the functions $f_a(\vec{\phi})$, $a = 1,2,3,$ depend on the components of the matrices $\Phi^m$ and give rise to the potentials appearing in the Hamiltonian. 
These functions are given by

\begin{align}
& f_1 (\vec{\phi}) = (\phi^1_1 \phi^2_2 - \phi^1_2 \phi^2_1), \\Ê\nonumber
& f_2 (\vec{\phi}) = (\phi^1_2 \phi^2_3 - \phi^1_3 \phi^2_2),  \\ \nonumber
& f_3 (\vec{\phi}) = (\phi^1_3 \phi^2_1 - \phi^1_1 \phi^2_3).
\end{align} 
   
\noindent We can then make the mentioned simplifications to find a model with a particular potential, or even easier, for a null potential. This last is the first one for which we will find a solution.

\section{Exact Solutions} 

Here we will show the explicit form of the ground state wave function for a few models using some specific configurations of the components of the $\phi^m_a$ variables.

\subsection{Solution 1}

The model is the following. We choose the bosonic variables to be (in units where $\hbar = 1$)
   
\begin{align}
& \phi^1_{1} = x, ~ \phi^1_{2} = \phi^1_{3} =  0,~~\phi^2_{1} = y, ~ \phi^2_{2} = \phi^2_{3} = 0, \\ \nonumber
& \pi^1_{1} = -i\frac{\partial}{\partial x},~~\pi^1_{j} = 0 ~~~j = 2,3, \\ \nonumber
& \pi^2_{1} = -i\frac{\partial}{\partial y},~~\pi^2_{j} = 0~~~j= 2,3, 
\end{align}

\noindent and as a consequence of this choice we have 

\begin{align} 
& f_1 (\vec{\phi}) = 0, \\Ê\nonumber
& f_2 (\vec{\phi}) =  0, \\ \nonumber
& f_3 (\vec{\phi}) =  0.
\end{align}

\noindent Hence, in this configuration there will be no bosonic potential and the $G_a$ operators take the simpler form

\begin{align}
& G_1 = - i (\Lambda_{21} \Lambda_{31} + \Lambda_{22} \Lambda_{32}), \\ \nonumber
& G_2 = - i (\Lambda_{31} \Lambda_{11} + \Lambda_{32} \Lambda_{12}), \\ \nonumber 
& G_3 = - i (\Lambda_{11} \Lambda_{21} + \Lambda_{12} \Lambda_{22}).
\end{align}

\noindent One should apply all the three $G_a$ operators to the wave function, however given eq. (\ref{salgebra}) and eq. (\ref{dirac2}) 
we have to consider only the operator $G_1$ acting on the wave function $| \Psi \rangle$, and consistently it is sufficient to apply also one of the supercharges, namely $Q_1 | \Psi \rangle = 0$. Relations between the eight components of $ | \Psi \rangle $ arise and the wave function as a row vector has the form $ | \Psi \rangle = (\psi_1,\psi_2,i\psi_1,i\psi_2,-i\psi_2,-i\psi_1,-\psi_2,-\psi_1)$, 
where we mean by this that, $\psi_3 = i\psi_1, \psi_4 = i\psi_2$, and so on. Because of the form of the differential equations for this two dimensional model, it is easier to solve them using polar coordinates $x = r \cos{\phi}, y = r \sin{\phi}$. The exact solution for the two independent components of the wave function results in

\begin{align}
& \psi_1 = \pm a r^{\kappa}e^{i\kappa \phi} \pm b r^{-\kappa}e^{-i\kappa \phi} , \\ \nonumber
& \psi_2 = \mp a r^{\kappa}e^{i\kappa \phi} \mp b r^{-\kappa}e^{-i\kappa \phi} .
\end{align}

\noindent Here $\kappa$ is a separation constant and $a,b$ are arbitrary constants and we can choose them to have a normalizable solution.

\subsection{Solution 2}

The following model is chosen in such a way that the bosonic part of the potential in its corresponding Hamiltonian takes the form of a one dimensional harmonic oscillator, this is  

\begin{align}
& \phi^1_{1} = x,~\phi^2_{2} = c = \text{constant}, \\ \nonumber
& \phi^1_{2} = \phi^1_{3} = \phi^2_{1} = \phi^2_{3} = 0 , \\ \nonumber
&  f_1 (\vec{\phi}) = cx , \\Ê\nonumber
& f_2 (\vec{\phi}) = 0 , \\ \nonumber
& f_3 (\vec{\phi}) = 0.
\end{align}

\noindent Following the same procedure as in the previous solution, we will have two independent components, $\psi_1$ and $\psi_2$, for the wave function $| \Psi \rangle = (\psi_1,\psi_2,\psi_2,\psi_1,\psi_1,\psi_2,\psi_2,\psi_1)$. 
The explicit form of these two components is

\begin{align}
& \psi_1 = (a+b)e^{\frac{3cx^2}{2}} +(a-b)e^{-\frac{3cx^2}{2}}, \\ \nonumber
& \psi_2 = i(a+b)e^{\frac{3cx^2}{2}} - i(a-b)e^{-\frac{3cx^2}{2}},
\end{align}   

\noindent where $a,b$ are arbitrary constants, if we choose them  to be $a = -b$, then we get a normalizable wave function. 

\subsection{Solution 3}

The next model has also a null potential, but it has three independent bosonic variables

\begin{align}
& \phi^1_{1} = x,~\phi^1_{2} = y,~\phi^1_{3} = z ,  \\ \nonumber
& \phi^2_{1} = 0,~\phi^2_{2} = 0,~\phi^2_{3} = 0, \\ \nonumber \\ \nonumber
& f_1 (\vec{\phi}) = 0 , \\Ê\nonumber
& f_2 (\vec{\phi}) = 0 , \\ \nonumber
& f_3 (\vec{\phi}) = 0.
\end{align}

\noindent We can see from eq. (\ref{salgebra}) and eq. (\ref{dirac2}) that for this particular configuration it is not necessary to impose any $G_a |\psi \rangle = 0$ because $\phi^2_i = 0$, $i = 1,2,3$. We will try to find 
a solution to $Q_2 |\psi \rangle = 0$. This is possible and the eight independent components of the wave function reduce to only two, these are $\psi_1$ and $\psi_5$, and the wave function would take the 
following form $| \Psi \rangle = (\psi_1,\psi_1,\psi_1,\psi_1,\psi_5,-\psi_5,-\psi_5,\psi_5)$. Besides, there is a coupling between these two components, so they are also related. 
The form of the solution of $\psi_1$ is a linear combination of the exponential functions that come out from every particular choice of signs  ($\pm$) in the exponential factor $\exp[\pm \sqrt{a~}x \pm \sqrt{b~}y \pm i\sqrt{a+b~}z]$, 
where $a,b$ are constants and because of the coupling, $\psi_5$ results proportional to $\psi_1$, so $\psi_5 = C\psi_1$. The form of the constant $C$ depends on the particular form of $\psi_1$. For instance 
$\psi_1$ and $\psi_5$ could be the following 

\begin{align}
& \psi_1 = c_1e^{\sqrt{a}x + \sqrt{b}y + i\sqrt{a+b}z} + c_2e^{\sqrt{a}x + \sqrt{b}y - i\sqrt{a+b}z}, \\ \nonumber
& \psi_5 = \left( \frac{ (c_1+c_2)(\sqrt{b} -i\sqrt{a})}{(c_1-c_2)\sqrt{a+b}} \right) \psi_1,
\end{align}

\noindent where $c_1,c_2$ are constants. We see that the constant $C$ that relates $\psi_1$ and $\psi_5$ depends on the particular choice of the constants $c_1,c_2$ in the linear combination of exponentials. 
The most general solution of $\psi_1$ can have eight terms, each one with a proportional constant $c_j, j = 1,2,...,8$ and the constant $C$ would depend on all these $c_j$.
 
\subsection{Solution 4}

This model has three bosonic components, one is constant, with a potential different from zero. These are 

\begin{align}
& \phi^1_{1} = x,~\phi^1_{2} = 0,~\phi^1_{3} = z , \\ \nonumber
& \phi^2_{1} = 0,~\phi^2_{2} = c,~\phi^2_{3} = 0 , \\ \nonumber
& f_1 (\vec{\phi}) = cx , ~~ \text{$c$ is a constant}, \\Ê\nonumber
& f_2 (\vec{\phi}) = -cz , \\ \nonumber
& f_3 (\vec{\phi}) = 0.
\end{align}

\noindent After imposing the only non-trivial condition $G_2 |\Psi \rangle = 0$ and solving the equations $Q_1 |\Psi \rangle = 0$ we have a wave function solution with just one independent component, $\psi_1$. The form of $\psi_1$ and its 
relation with the rest of the components is

\begin{align}
& \psi_1 = \exp{\left[ i\tan^{-1}\left( \frac{z}{x} \right)  \right]}, ~~~ \psi_1 = i\psi_2 \\ \nonumber
& \psi_1 = \psi_4 = \psi_5 = \psi_8. ~~~~ \psi_2 = \psi_3 = \psi_6 = \psi_7, 
\end{align}

\noindent and a constriction for this model arises, $x^2 + z^2 = \frac{1}{3c}$. 

We have seen so far that these reduced lower dimensional models have exact solutions. 
We also found that for some of the configurations we have used, the arbitrary constants in the components of the wave function solutions can be chosen so that they are normalizable. We could make an extension of the 
method to a higher dimensional model but, as the dimension of the matrix representation for $\Lambda_{a \alpha}$ grows exponentially with the space dimensions, it would be hard to handle the complete 
9-dimensional problem. It will be needed a powerful computational method to solve the complete matrix model, even for the SU($2$) group and even harder for a larger group. In the next section, we solve a particular 
4-dimensional model. Our intention is to show that the method can be directly extended to more dimensions and to any SU($N$) group even though, as we mention, these extended models will be difficult 
to handle.         

\section{SU(2) 4-dimensional extension}

In the 4-dimensional extension for the SU(2) group there are 12 bosonic variables $\phi^m_a$, $m=1,2,3,4.~~a= 1,2,3,$ and with $d = 4$ the number of supercharges is 4. The number of fermionic variables is also 12 and 
the dimension $M$ for the  matrix representation of these variables is $M = 64$. The $(64 \times 64)$ representation we choose for the twelve matrices $\Lambda_{a \alpha}$ that satisfies the algebra eq. (\ref{algebra}) is the following

\begin{align}
& \Lambda_{11} = (\pm1/\sqrt{2}) \sigma^1 \otimes 1 \otimes 1 \otimes 1 \otimes 1 \otimes 1, \label{mrep} \\ \nonumber
& \Lambda_{12} = (\pm1/\sqrt{2}) \sigma^2 \otimes 1 \otimes 1 \otimes 1 \otimes 1 \otimes 1, \\ \nonumber
& \Lambda_{13} = (\pm1/\sqrt{2}) \sigma^3 \otimes \sigma^1 \otimes 1 \otimes 1 \otimes 1 \otimes 1, \\ \nonumber
& \Lambda_{14} = (\pm1/\sqrt{2}) \sigma^1 \otimes \sigma^2 \otimes 1 \otimes 1 \otimes 1 \otimes 1, \\ \nonumber
& \Lambda_{21} = (\pm1/\sqrt{2}) \sigma^3 \otimes \sigma^3 \otimes \sigma^1 \otimes 1 \otimes 1 \otimes 1, \\ \nonumber
& \Lambda_{22} = (\pm1/\sqrt{2}) \sigma^3 \otimes \sigma^3 \otimes \sigma^2 \otimes 1 \otimes 1 \otimes 1, \\ \nonumber
& \Lambda_{23} = (\pm1/\sqrt{2}) \sigma^3 \otimes \sigma^3 \otimes \sigma^3 \otimes \sigma^1 \otimes 1 \otimes 1, \\ \nonumber
& \Lambda_{24} = (\pm1/\sqrt{2}) \sigma^3 \otimes \sigma^3 \otimes \sigma^3 \otimes \sigma^2 \otimes 1 \otimes 1, \\ \nonumber
& \Lambda_{31} = (\pm1/\sqrt{2}) \sigma^3 \otimes \sigma^3 \otimes \sigma^3 \otimes \sigma^3 \otimes \sigma^1 \otimes 1, \\ \nonumber
& \Lambda_{32} = (\pm1/\sqrt{2}) \sigma^3 \otimes \sigma^3 \otimes \sigma^3 \otimes \sigma^3 \otimes \sigma^2 \otimes 1, \\ \nonumber
& \Lambda_{33} = (\pm1/\sqrt{2}) \sigma^3 \otimes \sigma^3 \otimes \sigma^3 \otimes \sigma^3 \otimes \sigma^3 \otimes \sigma^1, \\ \nonumber
& \Lambda_{34} = (\pm1/\sqrt{2}) \sigma^3 \otimes \sigma^3 \otimes \sigma^3 \otimes \sigma^3 \otimes \sigma^3 \otimes \sigma^2, \\ \nonumber
\end{align}

\noindent where $\sigma^i, i = 1,2,3,$ are the Pauli matrices. The $(4 \times 4)$ Dirac representation for the $\Gamma^m$ matrices we use is
 
\begin{align}
& \Gamma^1 = \label{dirac4} \left(
\begin{array}{cccc}
0 & - I \\
- I & 0
\end{array} \right),~~
\Gamma^2 = \left(
\begin{array}{cccc}
0 & -i\sigma^2 \\
i\sigma^2 & 0
\end{array} \right), \\Ê\nonumber \\ \nonumber
& \Gamma^3 = \left(
\begin{array}{cccc}
I & 0 \\
0 & - I
\end{array} \right),~~
\Gamma^4 = \left(
\begin{array}{cccc}
0 & -i\sigma^3 \\
i\sigma^3 & 0
\end{array} \right).
\end{align}

\noindent In this case $I$ is the $2 \times 2$ identity matrix. Then, for example, the first supercharge takes the following form

\begin{align}
Q_1 = & -\Lambda_{13}\pi^1_1 - \Lambda_{23}\pi^1_2 - \Lambda_{33}\pi^1_3-\Lambda_{14}\pi^2_1 \label{QE1} \\ \nonumber
& -\Lambda_{24}\pi^2_2-\Lambda_{34}\pi^2_3 +\Lambda_{11}\pi^3_1+\Lambda_{21}\pi^3_2 \\ \nonumber
& +\Lambda_{31}\pi^3_3-i\Lambda_{13}\pi^4_1-i\Lambda_{23}\pi^4_2-i\Lambda_{33}\pi^4_3 \\ \nonumber
& -ig\Lambda_{11}(\phi^1_{2} \phi^4_{3} - \phi^1_{3} \phi^4_{2})-ig\Lambda_{21}(\phi^1_{3} \phi^4_{1} - \phi^1_{1} \phi^4_{3}) \\ \nonumber
& -ig\Lambda_{31}(\phi^1_{1} \phi^4_{2} - \phi^1_{2} \phi^4_{1}) -g\Lambda_{12}(\phi^1_{2} \phi^2_{3} - \phi^1_{3} \phi^2_{2}) \\ \nonumber
& -g\Lambda_{22}(\phi^1_{3} \phi^2_{1} - \phi^1_{1} \phi^2_{3})-g\Lambda_{32}(\phi^1_{1} \phi^2_{2} - \phi^1_{2} \phi^2_{1}) \\ \nonumber
& +ig\Lambda_{12}(\phi^2_{2} \phi^4_{3} - \phi^2_{3} \phi^4_{2})+ig\Lambda_{22}(\phi^2_{3} \phi^4_{1} - \phi^2_{1} \phi^4_{3}) \\ \nonumber
& +ig\Lambda_{32}(\phi^2_{1} \phi^4_{2} - \phi^2_{2} \phi^4_{1}) + g\Lambda_{13}(\phi^1_{2} \phi^3_{3} - \phi^1_{3} \phi^3_{2}) \\ \nonumber
& +g\Lambda_{23}(\phi^1_{3} \phi^3_{1} - \phi^1_{1} \phi^3_{3})+g\Lambda_{33}(\phi^1_{1} \phi^3_{2} - \phi^1_{2} \phi^3_{1}) \\ \nonumber
& -ig\Lambda_{13}(\phi^3_{2} \phi^4_{3} - \phi^3_{3} \phi^4_{2})-ig\Lambda_{23}(\phi^3_{3} \phi^4_{1} - \phi^3_{1} \phi^4_{3}) \\ \nonumber
& -ig\Lambda_{33}(\phi^3_{1} \phi^4_{2} - \phi^3_{2} \phi^4_{1}) + g\Lambda_{14}(\phi^2_{2} \phi^3_{3} - \phi^2_{3} \phi^3_{2}) \\ \nonumber
& +g\Lambda_{24}(\phi^2_{3} \phi^3_{1} - \phi^2_{1} \phi^3_{3})+g\Lambda_{34}(\phi^2_{1} \phi^3_{3} - \phi^2_{2} \phi^3_{1}).
\end{align}

\noindent In this extension we have a richer structure and  we can search for more interesting potentials. We notice that the search for the ground state solution is much more complicated because we have to solve 
systems of 64 coupled partial differential equations for the 64 components of the wave function. In this extension we will also restrict ourselves to solve some reduced models with particular assumptions on the 
bosonic degrees of freedom.

\section{Solutions to the extended model}

Using the matricial representation in eq. (\ref{mrep}) 
we need to solve, again, systems of matricial partial differential equations. In this case, every supercharge and gauge symmetry operator is a 
$(64 \times 64)$ operator matrix acting on a $64$ component wave function $| \Psi \rangle$. 

As an example we begin with the following model. We select two variables to be different from zero, $\phi^1_1 = x$, $\phi^2_2 = y$. In this particular model, we see from eq. (\ref{salgebra}) and eq. (\ref{dirac4}) that it is not necessary to impose the restrictions $G_a |\Psi \rangle = 0$ for any $G_a$. We need then to solve, for example,  $Q_1|\Psi \rangle = 0.$ As in the previous models the components of the wave function can be identified and 
one gets a solution with just two independent components $\psi_1$, and $\psi_2$. The independent components $(\psi_1,\psi_2)$ appear finally coupled in only two differential equations  

\begin{align}
& i\frac{\partial \psi_2}{\partial x} + \frac{\partial \psi_2}{\partial y} - 3xy\psi_1 = 0, \label{Nic} \\ \nonumber
& i\frac{\partial \psi_1}{\partial x} - \frac{\partial \psi_1}{\partial y} + 3xy\psi_2 = 0.
\end{align}

\noindent These two coupled components $(\psi_1,\psi_2)$ are related to other six components of the wave function in the following manner

\begin{align}
& \psi_8 =  i\psi_1, \psi_7 = i\psi_2, \psi_{19} = -i\psi_2,  \label{relations} \\ \nonumber
&  \psi_{20} = i\psi_1, \psi_{21} = -\psi_1, \psi_{22} = \psi_2. 
\end{align}

\noindent There are also other pairs of components coupled in the same way, here we write all these pairs of coupled components 

\begin{align}
& (\psi_1,\psi_2), (\psi_3, \psi_4),  (\psi_9, \psi_{10}), (\psi_{11}, \psi_{12}),  \\ \nonumber
& (\psi_{33},\psi_{34}),(\psi_{35},\psi_{36}),(\psi_{41},\psi_{42}),(\psi_{43},\psi_{44}).
\end{align}

\noindent If we identify all these pairs as one, then all the rest of the components are related to the only independent pair $(\psi_1,\psi_2)$ in a similar way as in eq. (\ref{relations}) and the wave function has only two 
independent components. This, now exact, reduced model obtained following our general procedure would correspond to the only two component one guessed in \cite{N3} to analyze the spectrum of the Hamiltonian 
eq. (\ref{Ham}). Our quantization approach leads to similar features. It allows however an exact solution which satisfies also the Hamiltonian operator including the fermionic sector at once. A continuum spectrum of the Hamiltonian is 
an expected feature in the connection between the supermembrane and D0-branes \cite{1.1,N3}. For several possible choices of the bosonic variables one gets trivial solutions. Here we show solutions to some non-trivial models.
  
\subsection{Solution 1}

Another choice of a model with three bosonic components is the following

\begin{align}
\phi^1_2 = x,~~\phi^2_2 = y,~~\phi^3_3 = c = \text{constant}
\end{align}

\noindent with all other bosonic variables equal to zero. In this case, we should impose the condition $G_3 | \Psi \rangle = 0$, with the consequence that many components of the wave function vanish. 
The only components that are different from zero are 

\begin{align}
& \psi_1, \psi_2, \psi_3, \psi_4, \psi_{21}, \psi_{22}, \psi_{23}, \psi_{24}, \\ \nonumber
& \psi_{41}, \psi_{42}, \psi_{43}, \psi_{44}, \psi_{61}, \psi_{62}, \psi_{63}, \psi_{64},
\end{align}

\noindent and solving $Q_1 |\Psi \rangle = 0$, leads to the solution which is in this case 

\begin{align}
& \psi_1 = A\exp{\left[ \frac{3c}{2}(x^2-y^2)\right]},~~\psi_{21} = -iA\exp{\left[ \frac{3c}{2}(x^2-y^2) \right]},  \\ \nonumber
& \psi_1 = \psi_{2} = \psi_3 = \psi_4 = \psi_{41} = \psi_{42} = \psi_{43} = \psi_{44} \\ \nonumber
& \psi_{21} = \psi_{22} = \psi_{23} = \psi_{24} =  -\psi_{61} = -\psi_{62} = -\psi_{63} = -\psi_{64}.
\end{align}

\noindent with $A$ a real constant.

\subsection{Solution 2}

The next one dimensional model has a more interesting form, it has four bosonic components and three of them are constants. The choice of variables is 

\begin{align}
& \phi^4_2 = x, ~~~~\phi^1_2 = \phi^4_3 = c,~~~\phi^1_3 = d, \\ \nonumber
\end{align}

\noindent where $c$ and $d$ are constants and the rest of the bosonic degrees of freedom are set to zero. The conditions $G_a |\Psi \rangle = 0$ are not required and the solution to $Q_1 | \Psi \rangle = 0$ 
leads to pairs of coupled components, for instance $(\psi_1,\psi_{37})$ that can be placed in a two dimensional base of two eigenvectors in the form

\begin{align}
\left( 
\begin{array}{ccc}
\psi_1 \\
\psi_{37}
\end{array} \right) & = \label{solext2} c_1\left(
\begin{array}{ccc}
1 \\
-i
\end{array} \right) \exp{\left[ -3 \left( c^2x-\frac{dx^2}{2} \right) \right] } \\ \nonumber
& ~~ +
c_2\left(
\begin{array}{ccc}
1 \\
i
\end{array} \right) \exp{\left[ 3 \left( c^2x-\frac{dx^2}{2} \right) \right] }.
\end{align} 

\noindent The rest of the components are also coupled in the same way and give the same solution, these pairs of coupled components are 

\begin{align*}
& (\psi_1,\psi_{37}),~~(\psi_2,\psi_{38}),~~(\psi_3,\psi_{39}),~~(\psi_4,\psi_{40}), \\ \noindent
& (\psi_5,\psi_{33}),~~(\psi_6,\psi_{34}),~~(\psi_7,\psi_{35}),~~(\psi_8,\psi_{36}), \\ \noindent
& (\psi_9,\psi_{45}),~~(\psi_{10},\psi_{46}),~~(\psi_{11},\psi_{47}),~~(\psi_{12},\psi_{48}), \\ \noindent
& (\psi_{13},\psi_{41}),~~(\psi_{14},\psi_{42}),~~(\psi_{15},\psi_{43}),~~(\psi_{16},\psi_{44}), \\ \noindent
& (\psi_{17},\psi_{53}),~~(\psi_{18},\psi_{54}),~~(\psi_{19},\psi_{55}),~~(\psi_{20},\psi_{56}), \\ \noindent
& (\psi_{21},\psi_{49}),~~(\psi_{22},\psi_{50}),~~(\psi_{23},\psi_{51}),~~(\psi_{24},\psi_{52}), \\ \noindent
& (\psi_{25},\psi_{61}),~~(\psi_{26},\psi_{62}),~~(\psi_{27},\psi_{63}),~~(\psi_{28},\psi_{64}), \\ \noindent
& (\psi_{29},\psi_{57}),~~(\psi_{30},\psi_{58}),~~(\psi_{31},\psi_{59}),~~(\psi_{32},\psi_{60}). \\ \noindent
\end{align*}

\noindent It can be seen that there are only two independent eigenvector solutions, one of them is normalizable and the other one is not. In the next section we will see that this last solution can be interpreted as the supersymmetric quantum solution of a cosmological model.

\section{Connection with SUSY Quantum Cosmology}

As it was discussed in the introduction, M(atrix) theory is related with the theory underlying the five consistent theories of strings and 11-dimensional supergravity, namely M-theory. It is then expected that the 
matrix models would be related to gravity as well. A relation between gravity and matrix models is that these models describe the properties of Schwarzschild black holes after compactifications on torus of various 
dimensions \cite{16.1,16.2,16.3,16.4,Bht1,Bht2,Bht3,Bht4}. It has also been proposed that one can construct a cosmological model corresponding to matrix theory, in \cite{Alvarez} the authors study the relation between 
this matrix model and Newtonian cosmology. In \cite{Gibbons}, using a homothetic ansatz, the authors were able to get the Friedmann equations from the classical 
equations of matrix theory. Here, it is important to remark that in all these cosmological models, the fermionic sector is absent and that they have been worked out only with classical matrix equations. We are dealing with the 
quantization of the matrix model of the supermembrane directly related to the matrix model of D0-branes. Then we can expect that, given that we are working with the quantization of a supersymmetric model, our 
solutions would be related to the solutions that have been obtained in different approaches to SUSY quantum cosmology \cite{jefe1,jefe3,jefe4,jefe6,Torres}. 
As it was pointed in \cite{jefe4}, there are several ways to carry out the quantization of a supersymmetric cosmological model. One of this options is to use the superfield approach, generalizing the minisuperspace of the 
FRW cosmological model (and other cosmological models of interest) introducing the corresponding superpartner coordinates of time $t \rightarrow (t,\eta,\bar{\eta})$ and working with the supersymmetric generalization 
of the FRW action in terms of the superfields corresponding to the scale factor and the lapse function respectively $R(t) \rightarrow \mathbb{R}(t,\eta,\bar{\eta})$ and 
$N(t) \rightarrow  \mathbb{N}(t,\eta,\bar{\eta})$ \cite{jefe6,Torres}.  We can actually note that one of our solutions resembles the supersymmetric cosmological solution found in \cite{Torres} with this last approach.

From the supercharges and the SU(2) symmetry operators we can calculate the Hamiltonian of every model we have solved using eq. (\ref{salgebra}). For the solutions in the extended 4-dimensional model 
the Hamiltonians are $(64 \times 64)$ non-diagonal 
matrix operators whose diagonal terms are related with its bosonic part. Let us first take a look at the second solution of our extended model eq. (\ref{solext2}) but identify now the variables and constants in this case as 

\begin{align}
\phi^4_2 = R(t),~~~\phi^1_2 = \phi^4_3 = \sqrt{\frac{Mc}{3}},~~~\phi^1_3 = \frac{\sqrt{\kappa}c^3}{ 3 G},
\end{align}

\noindent where $\kappa$ is a constant, $c$ is the speed of light and $G$ is the gravitational constant. We will also recover for this solution the constant $\hbar$. The bosonic (diagonal) part of the Hamiltonian is 

\begin{equation}
H_{ii} = p_R^2 + \frac{\kappa c^6}{G^2}R^2 - \frac{2Mc^4 \sqrt{\kappa}}{G}R + M^2c^2, ~~~ i=1,2,...,64 \label{hii}
\end{equation}

\noindent and the solution in eq. (\ref{solext2}) for the coupled $(\psi_1,\psi_{37})$ components of the wave function become

\begin{align}
\left( 
\begin{array}{ccc}
\psi_1 \\
\psi_{37}
\end{array} \right) & =  c_1\left(
\begin{array}{ccc}
1 \\
-i
\end{array} \right) \exp{\left[-\left( \frac{Mc}{\hbar}R-\frac{\sqrt{\kappa}c^3}{2 G \hbar}R^2 \right) \right]} \label{sol} \\ \nonumber
& ~~ +
c_2\left(
\begin{array}{ccc}
1 \\
i
\end{array} \right) \exp{\left[ \left( \frac{Mc}{\hbar}R-\frac{\sqrt{\kappa}c^3}{2 G \hbar}R^2 \right) \right] }. 
\end{align} 

Now we review how this solution arised in the context of SUSY quantum cosmology. Following \cite{jefe6,Torres} we rewrite here the supersymmetric action of the FRW cosmological model with a perfect fluid of 
barotropic equation of state $p = \gamma \rho$. For the case of dust $\gamma = 0$ and null cosmological constant $\Lambda = 0$  such action is given by

\begin{equation}
S = \int \left[  -\frac{c^2}{2NG}R \left( \frac{dR}{dt} \right)^2  + N\frac{\kappa c^4}{2 G}R - NMc^2  \right]dt \label{nonsusyaction}
\end{equation}

\noindent where $\kappa = -1,1,0$ is the curvature constant. $N(t)$ and  $R(t)$ are the lapse function and the scale factor respectively and $M$ is the mass parameter of dust. 
This action as given is invariant under reparametrizations of time $[t \rightarrow t+a(t)]$ when $N(t),R(t)$ transform in the following way

\begin{align}
\delta R = a \frac{dR}{dt},~~~~\delta N = \frac{d(aN)}{dt}.
\end{align}  

\noindent In the superfield formulation one introduces the superpartner coordinates of time in the superspace $t \rightarrow (t,\eta,\bar{\eta})$. The superspace valued Taylor series expansion of the superfields is

\begin{align}
& \mathbb{N}(t,\eta,\bar{\eta}) = N(t) + i\eta \bar{\psi} (t) + i\bar{\eta}\psi (t) + V(t)\eta \bar{\eta}, \\ \nonumber
& \mathbb{R}(t,\eta,\bar{\eta}) = R(t) + i\eta \bar{\lambda} (t) + i\bar{\eta}\lambda (t) + B(t)\eta \bar{\eta},
\end{align}

\noindent where $\psi (t),\bar{\psi}(t)$ are the complex gravitino, and $V(t)$ is a $U(1)$ gauge field. $B(t)$ in the superfield $\mathbb{R} (t,\eta ,\bar{\eta})$ is an auxiliary degree of freedom and $\lambda (t),\bar{\lambda}(t)$ 
are the fermionic partners of the scale factor $R(t)$. The supersymmetric generalization of eq. (\ref{nonsusyaction}) is 

\begin{equation}
S = \int \left[  -\frac{c^2}{2 G}\mathbb{N}^{-1}\mathbb{R}D_{\bar{\eta}}\mathbb{R}D_{\eta}\mathbb{R} + \frac{c^3 \sqrt{\kappa}}{2 G}\mathbb{R}^2 - Mc\mathbb{R}  \right] d\eta d \bar{\eta} dt, 
\end{equation} 

\noindent where $D_{\bar{\eta}} = -\frac{\partial}{\partial \bar{\eta}} - i\eta \frac{\partial}{\partial t}$ and $D_{\eta} = \frac{\partial}{\partial \eta} + i\bar{\eta}\frac{\partial}{\partial t}$ are the supercovariant derivatives of the global 
``small" supersymmetry of the generalized parameter corresponding to the time variable $t$. When the canonical formalism is applied to the supersymmetric Lagrangian it was shown \cite{jefe6,Torres} that the classical total 
Hamiltonian is given by 

\begin{equation}
H_c = NH + \frac{i}{2}\bar{\psi}S-\frac{i}{2}\psi \bar{S} + \frac{1}{2}VF,
\end{equation}

\noindent where $H$ is the Hamiltonian of the system, $S$ and $\bar{S}$ are the supercharges and $F$ is a $U(1)$ rotation generator. $N$, $\psi$, $\bar{\psi}$ and $V$ are now Lagrange multipliers. $H, S, \bar{S}$ 
and $F$ are first class constraints, hence the physical state wave functions $\Psi (R)$ are obtained from the conditions 

\begin{align}
H \Psi (R) = 0,~~ S \Psi (R),~~ \bar{S} \Psi (R) =0,~~ F \Psi (R) = 0. \label{WDW}
\end{align}

\noindent These constraints have a dependence on $R(t)$ and $\lambda (t),\bar{\lambda} (t)$. The explicit dependence of the Hamiltonian and supercharges on this variables is the following

\begin{align}
& H = -\frac{G}{2c^2R}\pi_R^2-\frac{\kappa c^4R}{2G}-\frac{M^2 G}{2R} + c^2 M\sqrt{\kappa} \label{const} \\ \nonumber
& - \frac{c \sqrt{\kappa}}{2R}\bar{\lambda}\lambda - \frac{M G}{2c R^2} \bar{\lambda}\lambda, \\ \nonumber
& S = \left( \frac{i G^{1/2}}{cR^{1/2}}\pi_R - \frac{c^2\sqrt{\kappa}R^{1/2}}{G^{1/2}} + \frac{M G^{1/2}}{R^{1/2}}  \right) \lambda, \\ \nonumber
& \bar{S} = \left( \frac{i G^{1/2}}{cR^{1/2}}\pi_R + \frac{c^2\sqrt{\kappa}R^{1/2}}{G^{1/2}} - \frac{M G^{1/2}}{R^{1/2}} \right) \bar{\lambda}, \\ \nonumber
\end{align}   

 \noindent where $\pi_{R}$ is the canonical momentum of the variable $R$. As a consequence of the superalgebra between the Hamiltonian and the supercharges, the physical states are found 
 by applying only the supercharge operators to the state wave function $\Psi (R)$. Using the following matrix representation for the fermionic variables  

\begin{align}
\lambda = \sqrt{\hbar}\sigma_{-}, ~~\bar{\lambda} = - \sqrt{\hbar}\sigma_{+},~~~
\sigma_{\pm} = \frac{1}{2}(\sigma^1\pm \sigma^2),
\end{align}

\noindent $\Psi (R)$ becomes a two component wave function $\Psi (R) = (\Psi_1,\Psi_2)$. The corresponding superquantum solutions \cite{Torres} for these two components are   

\begin{align}
& \Psi_1 = C\exp{ \left[ \left( \frac{Mc}{\hbar}R -\frac{\sqrt{\kappa}}{2 G \hbar}R^2 \right) \right] }, \label{sol1} \\ \nonumber 
& \Psi_2 = \tilde{C}\exp{ \left[ -\left( \frac{Mc}{\hbar}R- \frac{\sqrt{\kappa}}{2 G \hbar}R^2 \right) \right] }.  
\end{align}


\noindent We can see that our solution eq.  (\ref{sol}) and this one obtained in the superfield approach eq. (\ref{sol1}), are indeed the same. The diagonal part of the matrix Hamiltonian for every matrix model is associated 
to the bosonic sector of the system. If we want to find the solution to $H \Psi (R) = 0$, we can rewrite the Hamiltonian constraint in eq. (\ref{const}) and see that the bosonic part of this Hamiltonian, which is the one that 
does not depend on the fermionic variables ($\lambda,\bar{\lambda}$) is the same as the diagonal part of the matrix Hamiltonian in eq. (\ref{hii}) of our particular matrix model. We have then found the same solution in 
SUSY quantum cosmology and in matrix theory following our general procedure to solve matrix theory. 

The behavior of the $\Psi_1$ component is the expected when $R \rightarrow R_{sup}$, where $R_{sup}$ is the maximum radius of the universe. One of the components converges as $R \rightarrow \infty$ while the other 
one diverges, but the convergent solution can be isolated with the right choice of the constants. The scalar product of the $\Psi_1$ component in the measure $R^{1/2}dR$ is normalizable 
 
\begin{align}
1 & = C^2\int_0^{R_{sup}}{ \Psi_1^2R^{1/2}dR },  \\ \nonumber
& = C^2\int_0^{R_{sup}} \left(  e^{2MR -\frac{\sqrt{\kappa}}{\tilde{G}}R^2} \right) R^{1/2}dR, 
\end{align}

\noindent and in both cases the norm of the wave function can be defined with the inner product

\begin{align}
\langle \overline{\Psi} | \overline{\Psi} \rangle = \int_0^{R_{sup}}{ \overline{\Psi}^{\dagger} \overline{\Psi}R^{1/2}dR } . 
\end{align}

\noindent This review allowed us to know how to get a supersymmetric quantum solution for a particular cosmological model in the superfield approach, and we were able to relate it with 
ground state solutions we have got in the matrix model, in particular for the second solution of the extended model. It was then important to show in some detail how these supersymmetric quantum solutions 
are obtained and how they are related with possibly normalizable solutions in matrix theory. We want to remark,  after we showed the coincidence of the solutions, that this makes evident certain connection between 
solutions in matrix theory and solutions in cosmological models arising from supergravity.

\section{Conclusions}

We have considered finite $N$ matrix models, the relevant operators are the supercharges and those corresponding to the SU($N$) symmetry. The Hamiltonian operator is a consequence of them. 
The fermionic degrees of freedom are represented as Dirac-like gamma matrices \cite{jefe1,jefe4}. By means of this, we were able to propose a general procedure to find exact ground state solutions to 
SU($N$) matrix models. We have shown here some supersymmetric reduced models and found their exact ground state solutions being some of them normalizable. 
The method we propose allows the possibility to find exact solutions by means of the constraints related to the SU($N$) invariance and the corresponding supercharges, containing both 
the bosonic and the fermionic sectors. Even though we have solved reduced models, the extension to any dimension and to any SU($N$) gauge invariant group is possible following the proposed procedure. 
One could then search for exact solutions to the full 9-dimensional matrix model and for any finite $N$ although powerful computational tools would be 
needed because the dimensionality of the matrices involved make the problem hard to handle. Given the structure of the supercharges for all the allowed dimensions as functions of the 
momenta and the potentials of the variables involved, we expect that some solutions to the lower dimensional models would also be solutions to certain higher dimensional ones. This is the case 
for the solutions we have shown here. Some of the solutions found to the 2-dimensional model are also solutions to the 4-dimensional extension. A reduced model guessed and investigated in \cite{N3} 
results as an exact sub-model in our formalism. A particular interesting result is a relation between one of our solutions in the extended model with supersymmetric quantum cosmology. 
This guarantees that the reduced matrix model is supersymmetric. These relations reinforce the method of quantization we propose in this work for finite $N$ matrix theory \cite{1.6}. 
Besides, we encounter possible relations between several topics connected with our work; on the one hand, according to our 
results, there is a connection between reduced matrix theory models and SUSY quantum cosmology, also, already in \cite{Gibbons} the FRW cosmological models have been related with 
classical bosonic matrix theory. Even Schwarzschild black holes \cite{16.1,16.2,16.3,16.4} and their thermodynamical properties \cite{Bht1,Bht2,Bht3,Bht4} have been obtained by compactifications in matrix theory. 
On the other hand, in a recent work \cite{Damour} the authors show evidence of the conjecture relating supergravity and the dynamics of  a spinning particle moving in an infinite coset space by means of the quantum 
dynamics of the supersymmetric Bianchi IX cosmological model \cite{jefe4} and the operational structure of the constraints revealed a hidden hyperbolic Kac-Moody structure. It would be of interest to be able to 
identify reduced matrix model solutions associated to the Bianchi SUSY quantum cosmological models, in particular the Bianchi IX \cite{jefe4,Damour}, and by means of this try to be able to find a relation 
between the Kac-Moody structure and matrix theory. An analysis of our work regarding several of these possible connections is the matter for future research and is beyond the scope of this work.  
    
\acknowledgments{ \noindent We want to thank H. Nicolai for useful comments on his own work and H. Garc\'ia Compe\'an, R. Cordero and O. Loaiza-Brito for discussions on the Hamiltonian formulation 
and different approaches of quantization in field and string theory. 
This work was partially supported by PROMEP and CONACYT Grant 135023. J. L. L\'opez was supported by CONACYT Grant 43683.}

\bibliographystyle{unsrt}
\bibliography{matrix_2}

\end{document}